\documentstyle[12pt,psfig,amsfonts,amssymb]{elsart}

\begin{document}
\begin{frontmatter}
    
\title{Covariant axial form factor of the nucleon in a chiral
       constituent quark model}

\author[Graz,Pavia]{L.Ya. Glozman},
\author[Pavia]{M.~Radici},
\author[Graz]{R.F.~Wagenbrunn}, 
\author[Pavia]{S.~Boffi}, 
\author[Iowa]{W.~Klink}, and
\author[Graz]{W.~Plessas} 

\address[Graz]{Institut f\"ur Theoretische Physik, Universit\"at Graz, \\
Universit\"atsplatz 5, A-8010 Graz, Austria}

\address[Pavia]{Dipartimento di Fisica Nucleare e Teorica,
Universit\`a di Pavia, \\ and Istituto Nazionale di Fisica Nucleare, 
Sezione di Pavia, I-27100 Pavia, Italy}
\address[Iowa]{Department of Physics and Astronomy, University of Iowa, \\
Iowa City, IA 52242, USA}

\date{\today}
        
\begin{abstract}

The axial form factor $G_{A}$ of the nucleon is investigated for the
Goldstone-boson-exchange constituent quark model using the point-form
approach to relativistic quantum mechanics. The results, being 
covariant, show large contributions from relativistic boost effects.
The predictions are obtained directly from the quark-model wave functions,
without any further input such as vertex or constituent-quark
form factors, and fall remarkably close to the available experimental data.

\vspace{.2cm}
\noindent
{\sl PACS\/}: 12.39-x; 13.10.+q; 14.20.Dh

\noindent
{\sl Keywords\/}: axial form factor, constituent quark model, point-form
relativistic quantum mechanics

\end{abstract} 

\end{frontmatter}

\section{Introduction}
\label{intro}

Low-energy hadron phenomena in the nonperturbative regime of quantum
chromodynamics (QCD) are suitably described in terms of effective
degrees of freedom within models incorporating the relevant properties of QCD.
In particular, the phenomenon of spontaneous breaking of chiral symmetry
(SB$\chi$S) is known to reduce the original 
$\rm SU(3)_L\times SU(3)_R$ symmetry of QCD to an $\rm SU(3)_V$ 
vector symmetry. As a first
consequence, the practically massless current quarks acquire a dynamical
mass related to the nonzero value of the quark condensate~\cite{NJL}. 
Such a dynamical mass can be viewed as the mass of quasiparticles,
which can be interpreted as the constituent quarks commonly used in quark 
models~\cite{Kokk}. Just recently, this dynamical-mass generation has been
established by lattice QCD calculations, with the result that the 
constituent mass approaches values of 300 - 400 MeV at small 
(Euclidean) momenta~\cite{latt}. 
Another important consequence of SB$\chi$S is that collective quark-antiquark
excitations can be identified with Goldstone-boson fields~\cite{NJL} and these 
Goldstone bosons should be coupled to constituent quarks~\cite{georgi}.
The latter should thus consistently be included as proper effective 
degrees of freedom in the formulation of constituent quark models 
(CQM). Consequently, Goldstone-boson exchange (GBE) becomes responsible for
mediating the (residual) interaction between constituent quarks.

A chiral quark model built up in this spirit for light-flavor baryons
was suggested in Ref.~\cite{GBE}. It was further elaborated and then parametrized 
in a semirelativistic
framework~\cite{Graz1}. This version of the GBE chiral quark model
is used in the present work. It relies on a three-quark Hamiltonian
containing a relativistic kinetic-energy operator and a linear 
confinement, whose strength is taken according to the
string tension of QCD. The hyperfine interaction of the constituent 
quarks is derived from GBE. It is realized by
the exchange of octet plus singlet pseudoscalar mesons, where only the
spin-spin components are taken into account. The inclusion of the other
force components as well as the consideration of possible vector and
scalar exchanges are under investigation~\cite{PANIC99}. 
However, the spin-spin part is the most important ingredient for the 
hyperfine interaction and indeed it already provides for a very reasonable 
description of the low-energy spectra of all light and strange baryons.
In particular, the specific spin-flavor dependence of the short-range part of
the GBE interaction produces the 
correct level orderings of the lowest positive- and negative-parity excitations
and thereby offers a convincing solution to a long-standing problem in 
baryon spectroscopy. 

Beyond spectroscopy, however, a constituent quark model should in 
addition also provide for the description of other
low-energy hadron phenomena, such as electromagnetic form 
factors and transitions, mesonic decays etc. The GBE 
constituent quark model, in the version of Ref. \cite{Graz1}, has 
recently been applied, e.g., in a first study of the nucleon 
electromagnetic form factors (including proton and neutron
charge radii and magnetic moments)~\cite{GrazPV}.
>From this investigation it has in particular 
turned out that a proper treatment of relativistic effects in the 
three-quark system is most essential. In the present work we report on 
a study of the nucleon axial form factor. Again it is an immediate 
demand to follow an approach that allows for
a strict observation of relativistic covariance. In order to reach 
this aim we have chosen to investigate the problem in the framework of 
relativistic Hamiltonian dynamics (i.e. Poincar\'e-invariant quantum 
mechanics).

Among the various forms of relativistic Hamiltonian dynamics that can be
considered in terms of unitary representations of the Poincar\'e group,
first discussed by Dirac~\cite{Dirac}, the point form~\cite{KP,Klink1} 
in particular offers some specific advantages.
Specifically, the interactions are contained only in the 
4-momentum operator $P^\mu$ that generates the space-time evolution
through the covariant equation

\begin{equation}    
P^\mu |\Psi\rangle = p^\mu |\Psi \rangle \, ,
\label{eq:eigen}
\end{equation}

where $|\Psi \rangle$ is an element of the Hilbert space (for a system 
with a fixed number of particles). Therefore, the
unitary representations $U(\Lambda)$ of Lorentz transformations $\Lambda$,
consisting of boosts and spatial rotations of the wave functions $|\Psi\rangle$, 
contain no interactions at all and remain purely kinematic. The
theory is thus manifestly covariant. Furthermore, the different $P^\mu$'s
commute with each other so that they can be diagonalized simultaneously.
Considering a three-body problem with constituent (quark) masses $m_{i}$
and individual 3-momenta $\pol k_{i}$, the interactions can be
introduced through the so-called Bakamjian-Thomas (BT) construction~\cite{BT},
by adding to the free mass operator
$M_{0}=\sqrt{P_{0}^\mu P_{0,\mu}}=
\sum_i \sqrt{\pol k_i^{\, 2} + m_i^2}$
an interaction part $M_{I}$ so that
$M=\sqrt{P^\mu P_\mu}=M_{0}+M_{I}$. Then also the 4-momentum operator 
gets split into a free part $P_{0}^{\mu}$ and an interaction part 
$P_{I}^{\mu}$:
 
\begin{equation}
P^\mu = P_0^\mu + P_I^\mu = M V^\mu = (M_0 + M_I) V^\mu.
\label{eq:bt}
\end{equation}

Here $V^\mu$ is the 4-velocity of the system, which is {\it not} modified 
by the interactions (i.e., $V^\mu=V_{0}^\mu$).
The mass operator $M$ with interactions must satisfy the following conditions

\begin{equation}
\left[ V^\mu , M \right] = 0, \qquad
U(\Lambda) M U^{-1} (\Lambda) = M
\label{eq:algebra}
\end{equation}

in order to fulfil the Poincar\'e algebra of the 4-momentum operators.
In the center-of-momentum frame of the three-body system, for which
$\pol P = \sum_i \pol k_i = 0$ and $V^\mu = (1,0,0,0)$,
the stationary part of Eq. (\ref{eq:eigen}) can be identified with the
eigenvalue problem solved in Ref.~\cite{Graz1} for the GBE Hamiltonian
$H=\sum_i \sqrt{\pol k_i^{\, 2} + m_i^2}+H_{I}=H_{0}+H_{I}$. Therefore,
the latter, even if including in $H_{I}$ a phenomenological confinement
and an instantaneous hyperfine interaction, can be interpreted as
a mass operator fulfilling all the necessary commutation
relations of the Poincar\'e group. The corresponding eigenfunctions
describe all possible states of the three-body system in the
center-of-momentum frame. 

Under an arbitrary Lorentz transformation, 
each quark spin gets rotated by a different Wigner rotation $R_{W_i}$, thus
preventing the definition of a total spin for the rotated state. It is useful
to introduce the so-called velocity states~\cite{Klink2} by applying a 
particular Lorentz boost $B(v)$ to the  
eigenfunctions in the center-of-momentum frame. This boost takes the
whole system from the rest frame to a four-velocity $v$ with new four-momenta 
$p_i = B(v) k_i$ for the individual quarks.
Now, under any Lorentz transformation $\Lambda$, each individual quark 
spin and orbital angular momentum in the velocity states is 
rotated by the same Wigner rotation $R_W= B^{-1}(\Lambda v) \Lambda B(v)$
so that it is always possible to define 
a total spin in the same way as in nonrelativistic quantum mechanics.
In practice, the point-form approach together with the use of velocity
states allows for an exact calculation of all necessary transformations
of the momentum dependences and of relativistic 
quark-spin rotations associated with proper Lorentz boosts of the 
three-quark wave functions.

In the case of electromagnetic reactions, the point-form electromagnetic 
current operator $J^\mu$ can be written in terms of
irreducible tensor operators under the strongly interacting Poincar\'e
group~\cite{Klink1}. Thus, e.g., the nucleon charge and magnetic form factors
can be obtained as reduced matrix elements of such an irreducible
tensor operator in
the Breit frame with the virtual photon momentum along the $\hat z$ axis, i.e. 
$q^\mu_B = (0,0,0,q)$. By using the eigenfunctions of the GBE 
Hamiltonian~\cite{Graz1} boosted to velocity states in the Breit frame, these 
form factors have been calculated in Ref.~\cite{GrazPV}, assuming a 
single-particle current operator for point-like quarks. This approach
corresponds to a relativistic impulse approximation but specifically in
the point form. It is conventionally called Point-Form Spectator
Approximation (PFSA). Without introducing any further adjustable 
parameters (such as vertex cut-offs or quark form factors)
the corresponding results have been found to fall remarkably close to
existing experimental data for all elastic observables, 
i.e. proton and neutron electric as well as magnetic form factors and 
charge radii as well as magnetic moments. 

Here, we report predictions of the nucleon axial form factor that, in contrast to the
electromagnetic case, connects the proton wave function to the neutron one and,
therefore, it represents a further test of the model wave functions. In the 
following Section we outline the calculation of axial current matrix 
elements in the point-form approach. In Section \ref{results} we present 
the results and in Section \ref{why} we discuss some of the main reasons why this 
approach appears so promising. We give a short summary in Section \ref{end}.

\section{The axial form factor in the point-form approach}
\label{pf}

The axial-current matrix element between the initial and final
nucleon states with 4-momenta $p$ and $p^{\prime}$, and spins $s$ and $s^{\prime}$,
respectively, is defined as

\begin{equation}
\langle p',s'|A^\mu_a|p,s\rangle = \overline{u}(p',s')\left[G_A(Q^2)\gamma^\mu +
 \frac{1}{2M} G_P(Q^2)({p'}^\mu-p^\mu)\right] \gamma_5 \textstyle{\frac{1}{2}} 
 \tau_a u(p,s),
\label{eq:ax_def}
\end{equation}

where $M$ is the nucleon mass, $q^\mu={p'}^\mu - p^\mu$, 
$Q^2=\pol q^{\,2}-\omega^2\ge 0$,
$\tau_a$ is the isospin matrix with Cartesian index, and $u(p,s)$
is the usual Dirac spinor.
Here, $G_A(Q^2)$ is the axial form factor and $G_P(Q^2)$ the
induced pseudoscalar form factor. In the Breit frame, with the momentum
transferred only along the $\hat z$ axis, we have

\begin{equation}
p_B^\mu = (E_B, 0, 0, -\textstyle{\frac{1}{2}} |\pol q |), \quad\  p_B^{\prime \, 
\mu} = (E_B, 0, 0, \textstyle{\frac{1}{2}} |\pol q |) \, , \quad\ 
E_B = \sqrt{M^2+\textstyle{\frac{1}{4}} \pol q^{\, 2}} \, . 
\label{eq:breit}
\end{equation}

Therefore, one obtains

\begin{eqnarray}
\langle p'_B,s'|A^0_a|p_B,s\rangle & = & 0,\nonumber\\
\langle p'_B,s'|\pol A_a |p_B,s\rangle & = &
\chi^\dagger_{s'}\left[\frac{E_B}{M}G_A(Q^2){\pol \sigma}_T \right.\nonumber\\
& & \qquad\qquad + \left. \left(G_A(Q^2) -
\frac{\pol q^{\, 2}}{4M^2}G_P(Q^2)\right)\pol \sigma_L\right]
\textstyle{\frac{1}{2}} \tau_a\chi_s \, ,
\label{eq:ga}
\end{eqnarray}

where

\begin{equation}
{\pol \sigma}_T = \pol \sigma - {\hat q}({\pol \sigma}\cdot{\hat q}) \, ,
\qquad {\pol \sigma}_L = {\hat q}({\pol \sigma}\cdot{\hat q})
\end{equation}

and $\chi_s$ is the two-component spinor of the nucleon.

The axial form factor $G_A(Q^2)$ is the only contribution to the transverse part
of the current that is not affected by current conservation. 
Therefore, one can obtain $G_A$ by applying to $A_a^\mu$ the same PFSA 
approach as followed in Refs.~\cite{GrazPV,Klink1} for the 
electromagnetic current $J^\mu$. In
the following, we shall use for the initial nucleon state the neutron
wave function and for the final nucleon state the proton wave function.
Consequently, the isospin index $a$ in Eq. (\ref{eq:ga})
may be suppressed whenever not needed. In the Breit frame, the matrix elements of 
the transverse components of the axial current $A^{i =1,2}$ can be connected to the 
reduced matrix elements of the corresponding irreducible tensor of the Poincar\'e 
group, i.e.
  
\begin{equation}
\langle p'_B,s'|A^i |p_B,s \rangle  = G^i_{s's} \, , \qquad i=1,2 \, ,
\label{eq:zero}
\end{equation}

with the following identifications from Eq. (\ref{eq:ga}):

\begin{eqnarray}
G^1_{s's} & = & \frac{E_B}{M}G_A \chi^\dagger_{s'}\sigma_x\chi_s
=  \frac{E_B}{M}G_A(\delta_{s',s+1} + \delta_{s',s-1}) \, ,
\nonumber\\
G^2_{s's} & = & \frac{E_B}{M}G_A \chi^\dagger_{s'}\sigma_y\chi_s
=  -i \frac{E_B}{M}G_A(\delta_{s',s+1} - \delta_{s',s-1}) \, .
\label{eq:gaa}
\end{eqnarray}

The combined invariance under parity and time reversal gives

\begin{equation}
G^1  = \left( \begin{array}{cc}0 & G\\ G & 0\end{array}\right) \, , \quad
G^2 = -i\left( \begin{array}{cc}0 & G\\ -G & 0\end{array}\right) \, ,
\label{eq:quadue}
\end{equation}

with $G$ real. The comparison with Eq. (\ref{eq:gaa}) thus gives

\begin{equation}
\frac{E_B}{M}G_A = G \, .
\end{equation}

The calculation of the reduced matrix elements $G_{s's}^i$ in Eq.
(\ref{eq:zero}) is made in PFSA and follows the same
lines of the formalism developed in Ref.~\cite{Klink1} and already used in
Ref.~\cite{GrazPV}.  Then one has

\begin{eqnarray}
\label{eq:invff}
G_{s's}^i(Q^2)&=&3\int d\pol k_1 d\pol k_2 d\pol k_3 d\pol k'_1 
d\pol k'_2 d\pol k'_3 \, \delta (\pol k_1 + \pol k_2 + \pol k_3 ) \,
\delta (\pol k'_1 + \pol k'_2 + \pol k'_3 ) \nonumber \\
&&\hspace{-2em}\times 
\psi^\ast_{s'}(\pol k'_1,\pol k'_2,\pol k'_3;\mu'_1,\mu'_2,\mu'_3) \, \, 
 \psi_{s}(\pol k_1,\pol k_2,\pol k_3; \mu_1,\mu_2,\mu_3) \nonumber\\
&&\hspace{-2em}\times{D^{1/2}_{\lambda'_1\mu'_1}}^\ast[R_W(k'_1,B(v_{\rm out}))]
\langle p'_1,\lambda'_1 |A^i_{[1]}|p_1,\lambda_1\rangle
D^{1/2}_{\lambda_1\mu_1}[R_W(k_1,B(v_{\rm in}))]\nonumber\\
&&\hspace{-2em}\times D^{1/2}_{\mu'_2\mu_2}[R_W(k_2,B^{-1}(v_{\rm out})
B(v_{\rm in}))] D^{1/2}_{\mu'_3\mu_3}[R_W(k_3,B^{-1}(v_{\rm out})B(v_{\rm in}))]
\nonumber\\
&&\hspace{-2em}\times\delta^3[k'_2-B^{-1}(v_{\rm out})B(v_{\rm in})k_2]
\delta^3[k'_3-B^{-1}(v_{\rm out})B(v_{\rm in})k_3] \, ,
\label{eq:dodi}
\end{eqnarray}

where a summation is understood for repeated indices and the initial and final 
4-velocities are $M v_{\rm in} = p_B$ and $M v_{\rm out} = p'_B$, respectively. 
In Eq.~(\ref{eq:dodi}) $\psi_{s}$ is the nucleon wave function in the 
centre-of-momentum frame with
$\pol k_i$ and $\mu_i$ being the individual quark momenta and spin
projections, respectively, and $D^{1/2}$ is the standard rotation matrix.
The single-quark axial-current matrix element has the form

\begin{equation}
\langle p'_i,\lambda'_i|A^\mu_{a\, [i]}|p_i,\lambda_i\rangle =
g_A^q \bar{u}(p'_i,\lambda'_i) \gamma^\mu \gamma_5 \textstyle{\frac{1}{2}} 
\tau_a  u(p_i,\lambda_i),
\label{eq:q_axial}
\end{equation}
where $u(p_i,\lambda_i)$ is the Dirac spinor of quark $i$ with momentum $p_i$
and spin
projection $\lambda_i$, and $\tilde{q}=p'_{i}-p_{i}$ 
is the momentum transferred to a single quark. The
quark axial charge is assumed to be $g_A^q=1$, as for free bare fermions. 

\begin{figure}[h]
\vspace*{-0.3cm}
\hspace*{0.5cm}\psfig{file=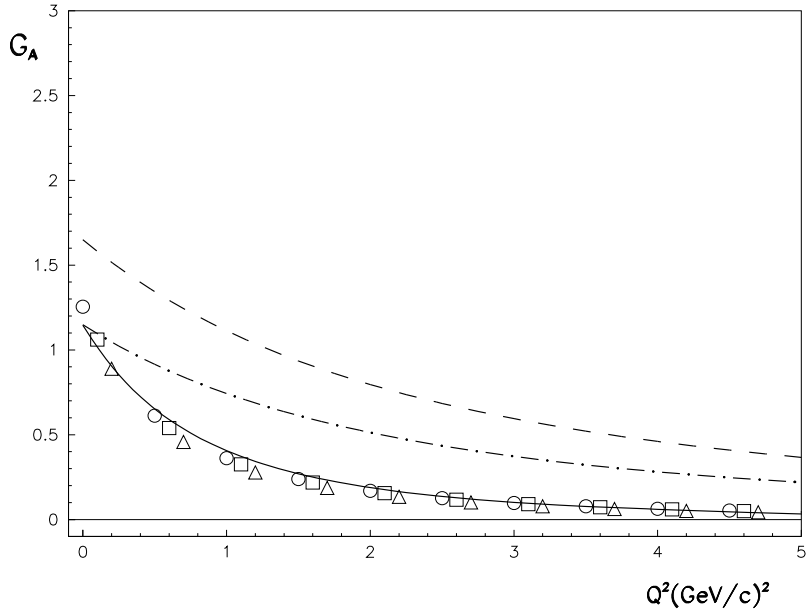,width=28pc}
\caption{Nucleon axial form factor $G_{A}$. Solid line: Fully 
relativistic PFSA result.
Dashed line: Nonrelativistic result. Dot-dashed line: Result with 
relativistic current operator but without boosts.
The experimental data are plotted following the dipole form of 
Eq.~(\ref{eq:dipole}). Squares and circles correspond to different 
values for the axial mass $M_A$ from charged-pion electroproduction experiments;
triangles correspond to the world average $M_A$ value from neutrino (antineutrino)
scattering on protons and nuclei (see the text).}
\label{fig:gamainz}
\end{figure}

\section{Results}
\label{results}

The prediction for the axial form factor, 
calculated in relativistic PFSA as described in the previous Section, is given
by the solid curve in Fig.~\ref{fig:gamainz}. For comparison,
the dashed line represents the purely nonrelativistic result obtained
if the nonrelativistic expression for the quark axial current
is adopted and no boosts are applied to either the initial or final nucleon
states. In this case, the axial constant is $g_A \equiv 
G_A(0) = 1.65$ and the marginal deviation from the value 
$\textstyle{\frac{5}{3}}$ predicted with SU(6) harmonic oscillator wave functions
is due to the small admixture of mixed-symmetry components in the nucleon wave
functions of the GBE quark model of Ref.~\cite{Graz1}. The dot-dashed line
in Fig.~\ref{fig:gamainz} shows the result if the relativistic quark axial
current is adopted but no boosts are applied (cf. Ref.~\cite{Riska}).
Comparing this result to the full line, it becomes evident that at $Q^2=0$
the boosts do not contribute and the axial constant adopts the same 
value as in the complete relativistic calculation. 

The experimental data are presented in Fig.~\ref{fig:gamainz} assuming
the dipole form

\begin{equation}
G_A(Q^2) = \frac{g_A}{\left( 1+ \displaystyle{\frac{Q^2}{M_A^2}} 
\right)^2},
\label{eq:dipole}
\end{equation}

where the axial constant was taken to be $g_A=1.255\pm0.006$, as obtained from 
$\beta$-decay experiments~\cite{fa}. For the axial mass $M_A$ we used 
three different values, namely the world average $M_A=1.069\pm0.016$ GeV 
from charged-pion electroproduction (represented by squares), the most recent 
value $M_A=1.077\pm0.039$ GeV from the p(e,e$'\pi^+$)n experiment at 
Mainz~\cite{mainz} (represented by circles), and the world average
$M_A=1.032\pm0.036$ GeV from neutrino (antineutrino) scattering experiments on
protons and nuclei~\cite{neutr} (represented by triangles).

At $Q^2=0$ the predicted value of the axial constant $g_A \equiv 
G_A(0)=1.15$ is a bit lower than the experimental one. It is not yet 
clear which effect is responsible for this behaviour. We can think 
of a number of reasons (e.g., a quark axial constant $g_A^q$ 
different from 1, etc.), which, however, would require a series of
further investigations going beyond the scope of this paper.

Similarly, the axial radius deduced from the slope of $G_{A}$
at $Q^2=0$ is also lower than the experimental one. 
For the GBE quark model of Ref. ~\cite{Graz1} one gets
$<r_A^2>^{1/2} = 0.520 \ \hbox{\rm fm}$, whereas the experimental value 
is $<r_A^2>^{1/2} = (0.635\pm 0.023)\ \hbox{\rm fm}$, as extracted from pion 
electroproduction~\cite{mainz}, or $<r_A^2>^{1/2} = (0.65\pm 0.07)\ \hbox{\rm fm}$,
as deduced from neutrino experiments~\cite{neutr}. 

All the results were calculated with the nucleon wave functions from 
the GBE quark model as the only input. Point-like constituent quarks 
were assumed and no further phenomenological parameters were 
introduced. For the $Q^2$-range shown in Fig.~\ref{fig:gamainz}, the 
PFSA results just fall on top of the experimental data. These results 
are similarly remarkable as before in the case of elastic
electromagnetic form factors, where the PFSA results 
also came quite close to the experimental data~\cite{GrazPV}.
The comparison with the nonrelativistic result (dashed line) shows a 
large discrepancy. On the one hand this is due to the use of a 
nonrelativistic current operator and on the other hand it misses the 
relativistic boost effects. That the latter are of considerable 
importance can be deduced from the comparison with the `intermediate' 
result represented by the dot-dashed curve (corresponding to the case 
with a relativistic current but no boosts). All this is completely
in line with previous results for the electromagnetic form
factors~\cite{GrazPV}.

\section{Discussion}
\label{why}

Considering the results for the axial form factor as presented in the 
previous Section, together also with the electromagnetic-form-factor 
results from a completely analogous study published in Ref.
\cite{GrazPV}, one may wonder why predictions are
obtained such that in all aspects they are readily in accordance
with the experimental data and thus produce a consistent
miscroscopic picture of the structure of the 
nucleons. This is especially remarkable in view of various previous 
attempts to explain the nucleon form factors (at low momentum 
transfers) from constituent quark models.

We argue for two main ingredients that are essential for achieving 
such results: firstly the strict observation of relativistic 
effects, secondly the usage of realistic quark model wave 
functions.

Certainly, confined few-quark systems have to be treated in a 
relativistic framework. For three-quark systems, such as the nucleons,
it has been quite difficult so far to fulfill this demand, especially 
with regard to generating the wave functions and calculating covariant 
observables. The reasons are manifold and cannot all be discussed 
here. 
We have found that relativistic Hamiltonian dynamics provides a 
promising approach to account for relativistic effects in confined 
few-constituent-quark systems. Specifically the point-form version is well adapted
for an exact calculation of the necessary boosts of the wave functions
and, consequently, for a strictly covariant calculation of matrix
elements for any observables. Poincar\'e-invariant quantum mechanics, 
in any of its formulations (such as point, instant, or front forms
\cite{Dirac}), is 
rigorously defined on a Hilbert space of a finite number of 
particles. In this respect it is most appropriate for a relativistic 
treatment of constituent quark models, which, as effective models of 
QCD for low-energy hadrons, are also built with a finite number of 
degrees of freedom. In such a situation, relativistic Hamiltonian 
dynamics for few-particle problems lacks only the property of cluster
separability~\cite{sokolov,coester,weinberg}. However, for few-quark systems, it is
legitimate to argue that this feature is not really important whenever 
the constituents cannot be separated asymptotically (as it is 
specifically the case 
for the nucleons as stable three-quark bound states).

The main characteristics of the GBE CQM have already been mentioned
in the Introduction and expressed in much detail in several places in 
the literature \cite{GBE,GlozPANIC}.  
With respect to the form factor results we emphasize only those 
properties that are essential. Obviously 
the baryon wave functions must have the correct spatial extensions and 
the required symmetries. The GBE CQM comes with a specific 
spin-flavor symmetry that arises from its theoretical foundation 
\cite{GBE} and is constrained by a fit to baryon spectroscopy.
The parametrization in Ref. \cite{Graz1} achieves a unified 
description of all light and strange baryon spectra in good agreement 
with phenomenological data. It turns out that also the eigenfunctions 
of the corresponding Hamiltonian are quite realistic. For the 
description of both the electromagnetic and axial structure of the
nucleons rather subtle properties of the wave functions are essential.
For instance, to simultaneously reproduce both the proton and 
neutron electromagnetic form factors in Ref. \cite{GrazPV}
as well as the axial form factor in the present work
both the overall structure and the rather small mixed-symmetry 
configurations in the nucleon wave functions are crucial. 
Otherwise one would not obtain a consistent description without 
introducing further parameters beyond the quark model.

It is satisfying that in addition to these experimental 
constraints the GBE CQM also meets well-established properties 
of QCD in the low-energy regime. It respects the important 
consequences of SB$\chi$S (adressed in the Introduction) and is 
also consistent with the large $N_c$ behaviour of QCD, where the SU(6)
symmetry of baryons becomes exact~\cite{last}. The particular confinement
interaction used in the GBE CQM
provides baryon wave functions with exactly this symmetry, while
the GBE hyperfine interaction gives rise to the breaking of the SU(6)
symmetry at lower order; for the nucleons, in particular, it 
produces the admixture of small mixed-symmetry components
at order $1/N_c$, which is among the most important ingredients
for the form factor results.

The present results have been obtained with a one-body current 
operator (in the point-form approach) only. In principle, one would
also expect contributions from 
two-body operators. If they were large, they would spoil the good 
results. However, contributions from exchange-currents are intimately
related to relativistic effects. In fact the PFSA used here is not the 
usual nonrelativistic impulse approximation, because the impulse
given to the nucleon is not the same as the impulse given to
the struck quark. Further, the one-body current is not only
covariant, in the electromagnetic case it is also conserved.
It remains to be seen 
by quantitative calculations how much possible remaining two-body currents
will contribute. Since relativistic effects are fully accounted for in 
the point-form approach, there is good hope that beyond the PFSA
any contributions from genuine two-body currents will remain small.

\section{Summary}
\label{end}

We have presented first results for the nucleon axial form 
factor predicted by the GBE constituent quark model of
Ref.~\cite{Graz1}. They were obtained in a covariant theory using 
the point-form approach of relativistic Hamiltonian dynamics.
The full PFSA results, including all relativistic effects, are
found to fall remarkably close to the experimental data at low and 
moderate momentum transfers without introducing any further parameters. This behaviour is 
in striking correspondence with the case of the electromagnetic observables
considered so far for the GBE quark model; in
Ref.~\cite{GrazPV} practically the same characteristics were found 
with regard to electromagnetic neutron as well as proton elastic form factors. 
It should be emphasized that the axial form factor also represents a 
stringent test for the quality of the nucleon wave functions. 
Again, as in the case of the electromagnetic observables, several 
realistic characteristics of the quark model wave functions, such as 
mixed-symmetry components or a proper size of the nucleons, are 
required to attain a reasonable result. The GBE quark model obviously 
provides such features for the baryon wave functions, in addition
to producing the correct eigenenergies in the excitation spectra.
All this indicates that the GBE interaction may be a reasonable phenomenological
representation of the low-energy strong interaction. 

In any case it has become obvious that relativistic effects are crucially 
important. Using Poincar\'e-invariant quantum mechanics appears
to be a promising way of including them in a definite
manner. In particular, the point-form approach makes it possible to reliably 
calculate relativistic effects for three-quark systems without the necessity 
of introducing any approximations.

\end{document}